\documentclass[twocolumn,pra,aps,superscriptaddress]{revtex4-1}
\usepackage{amssymb,amsmath,amsfonts}

\usepackage[latin9]{inputenc}
\usepackage{color}
\usepackage{amstext}
\usepackage{graphicx}
\usepackage[colorlinks=true,linkcolor=blue,urlcolor=blue,citecolor=blue,pdfusetitle]{hyperref}

\usepackage{hyperref,cleveref}
\usepackage[dvipsnames]{xcolor}
\usepackage[caption=false]{subfig}

\makeatletter

\newcommand\adag{a^\dagger}

\newcommand\braket[1]{\left\langle #1\right\rangle}

\usepackage{times}
\usepackage{bbm}

\makeatother

\begin{document}
\title{Assessing the role of initial correlations in the entropy production rate for nonequilibrium harmonic dynamics}
\date{\today}

\author{Giorgio Zicari}\email{gzicari01@qub.ac.uk}
\affiliation{Centre for Theoretical Atomic, Molecular, and Optical Physics, School of Mathematics and Physics, Queen's University, Belfast BT7 1NN, United Kingdom}

\author{Matteo Brunelli}
\affiliation{Cavendish Laboratory, University of Cambridge, Cambridge CB3 0HE, United Kingdom}

\author{Mauro Paternostro}
\affiliation{Centre for Theoretical Atomic, Molecular, and Optical Physics, School of Mathematics and Physics, Queen's University, Belfast BT7 1NN, United Kingdom}

\begin{abstract}

Entropy production provides a general way to state the second law of thermodynamics for non-equilibrium scenarios. In open quantum system dynamics, it also serves as a useful quantifier of the degree of irreversibility. In this work we shed light on the relation between correlations, initial preparation of the system and non-Markovianity by studying a system of two harmonic oscillators independently interacting with their local baths. Their dynamics, described by a time-local master equation, is solved to show --  both numerically and analytically -- that the global purity of the initial state of the system influences the behaviour of the entropy production rate and that the latter depends algebraically on the entanglement that characterises the initial state.
\end{abstract}

\maketitle

\section{Introduction}
Entropy production plays a fundamental role in both classical and quantum thermodynamics: by being related to the second law  at a fundamental level, it enables to identify and quantify the irreversibility of physical phenomena \cite{DeGrootMazur}. This intimate connection has raised a great deal of interest in relation to the theory of open quantum systems, where one is concerned about the dynamics of a system interacting with the infinitely many environmental degrees of freedom \cite{Breuer-Petruccione}. In this scenario, a plethora of genuine quantum effects is brought about and a general and exhausting theory of entropy production is hitherto missing \cite{Mauro_thermo2018}.

The second law of thermodynamics can be expressed in the form of a lower bound to the entropy change $\Delta S$ undergone by the state of a given system that exchanges an amount of heat $Q$ when interacting with a bath at temperature $T$, that is
\begin{align}
\Delta S \ge \int \frac{\delta Q}{T} \ .
\end{align}
The strict inequality holds if the process that the system is undergoing is irreversible. One can thus define the entropy production $\Sigma$ as
\begin{align}
\label{eq:entropy_prod_def}
\Sigma \equiv \Delta S - \int \frac{\delta Q}{T} \ge 0 \ .
\end{align}
From \Cref{eq:entropy_prod_def}, one can obtain the following expression involving the rates~\cite{Landi2013, PhysRevLett.118.220601}:
\begin{align}
\frac{{\rm{d}} S}{{\rm{d}} t} = \Pi(t) - \Phi(t) \ ,
\end{align}
where $\Pi(t)$ is the entropy production rate and $\Phi(t)$ is the entropy flux from the system to the environment: at any time $t$, in addition to the entropy that is flowing from the system to the environment, there might thus be a certain amount of entropy intrinsically produced by the process and quantified by $\Pi(t)$. 

Entropy production is an interesting quantity to monitor in the study of open quantum systems, since this is the context where irreversibility is unavoidably implied. The issue has been addressed in order to obtain an interesting characterisation and measure of the irreversibility of the system dynamics \cite{Mauro_thermo2018}. In particular, it has been recently shown that the entropy production of an open quantum system can be split into different contributions: one is classically related to population unbalances, while the other is a genuine quantum contribution due to coherences \cite{Landi:19, POLKOVNIKOV:2011, PhysRevE.99.042105}. This fundamental result holds whenever the system dynamics is either described by a map microscopically derived through the Davies approach or in the case of a finite map encompassing thermal operations~\cite{Landi:19}. 
Most of these works, though,  are solely focused on the Markovian case, when the information is monotonically flowing from the system to the environment. Under this hypothesis, the open dynamics is formally described by a quantum dynamical semigroup; this is essential to mathematically prove that the entropy production is a non-negative quantity \cite{Spohn1978, Alicki_1979, PhysRevA.68.032105}. 
Moreover, whenever the quantum system undergoing evolution is composite (i.e., multipartite) beside system-environment correlations, also inter-system correlations will contribute to the overall entropy production.  A full account of the role of such correlations (entanglement above all) on the entropy balance is not known.

However, a strictly Markovian description of the dynamics does not encompass all possible evolutions. There might be circumstances in which there is no clear separation of time-scales between system and environment: this hampers the application of the Born-Markov approximation \cite{Breuer-Petruccione}. In some cases, a backflow of information going from the environment to the system is observable, usually interpreted as a signature of a quantum non-Markovian process \cite{Rivas:14, Breuer:16}. From a thermodynamical perspective the non-negativity of the entropy production rate is not always guaranteed, as there might be intervals of time in which it attains negative values. It has been argued that this should not be interpreted as a violation of the second law of thermodynamics \cite{Benatti2017}, but it should call for a careful use of the theory, in the sense that -- in the entropy production balance -- the role of the environment cannot be totally neglected. This idea can be justified in terms of the backflow of information that quantum non-Markovianity entails: the system retrieves some of the information that has been previously lost because of its interaction with the surroundings.

In this paper,  we investigate the way initial correlations affect the entropy production rate in an open quantum system by considering the case of non-Markovian Brownian motion. 
We focus on the case of an uncoupled bipartite system connected to two independent baths. The rationale behind this choice is related to the fact that any interaction between the two oscillators would likely generate, during the evolution, quantum correlations between the two parties. In general, the entanglement dynamically generated through the interaction would be detrimental to the transparency of the picture we would like to deliver, as it would be difficult to isolate the contribution to $\Pi(t)$ coming from the initial inter-system correlations. To circumvent this issue, in our study we choose a configuration where the inter-system dynamics is trivial (two independent relaxation processes) but the bipartite state is initially correlated. \textit{De facto}, the entanglement initially present in the state of our ``medium'' acts as an extra knob which can be tuned to change the rate of entropy production, thus steering the thermodynamics of the open system that we consider.

The paper is organised as follows. In \Cref{sec:Gauss} we introduce a closed expression of the entropy production rate for a system whose dynamics is described in terms of a differential equation in the Lyapunov equation. In \Cref{sec:QBM} we introduce the model we would like to study: a system of two uncoupled harmonic oscillators, described by a non-Markovian time-local master equation. We also discuss the spectral properties of the two local reservoirs. This minimal, yet insightful, setting allows us to investigate -- both numerically and analytically -- how different initial states can affect the entropy production rate. We investigate this relation in depth in \Cref{sec:main}, where, by resorting to  a useful parametrisation for two-mode entangled states, we focus on the role of the purity of the total two-mode state and on the link between the entanglement we input in the initial state and the resulting entropy production. In \Cref{sec:Markovian_Limit} we assess whether our results survive when we take the Markovian limit. Finally, in \Cref{sec:conclusions}, we summarise the evidence we get and we eventually draw our conclusions.

\section{Entropy production rate for Gaussian systems}
\label{sec:Gauss}

We restrict our investigation to the relevant case of Gaussian systems.
\cite{Ferraro:05, Serafini:17, Carmichael}. This choice dramatically simplifies the study of our system dynamics, since the evolution equations only involve the finite-dimensional covariance matrix (CM) of the canonically conjugated quadrature operators. 
According to our notation, the CM $\boldsymbol{\sigma}$, defined as
\begin{align}
\label{eq:cov_def}
\sigma_{ij} =\braket{\{X_i, X_j \}} - 2\braket{X_i}\braket{X_j}  ,
\end{align}
satisfies the Lyapunov equation
\begin{align}
\label{eq:Lyapunov}
\dot{\boldsymbol{\sigma}} = \mathbf{A} \boldsymbol{\sigma} + \boldsymbol{\sigma} {\mathbf{A}^{\rm{T}}} + \mathbf{D} ,
\end{align}
where $\mathbf{A}$ and $\mathbf{D}$ are the drift and the diffusion matrices, respectively, and $\mathbf{X} = \{q_1,p_1,\ldots, q_N,p_N\}^{\rm T}$ is the vector of quadratures for $N$ bosonic modes. 
In particular, the CM representing a two-mode Gaussian state can always be brought in the standard form \cite{Ferraro:05, Serafini:17}:
\begin{align}
\label{eq:sigma_sf}
\boldsymbol{\sigma} = \begin{pmatrix}
a & 0 & c_{+} & 0 \\
0 & a & 0 & c_{-} \\
c_{+} & 0 & b & 0 \\
0 & c_{-} & 0 & b 
 \end{pmatrix},
\end{align}
where the entries $a, \ b$, and $c_{\pm}$ are real numbers. Furthermore, a necessary and sufficient condition for separability of a two-mode Gaussian state is given by the Simon criterion \cite{PhysRevA.72.032334}:
\begin{align}
\label{eq:uncertainty}
\tilde{\nu}_{-} \ge 1, 
\end{align}
where $\tilde{\nu}_{-}$ is the smallest symplectic eigenvalue of the partially transposed CM $\tilde{\boldsymbol{\sigma}} = \boldsymbol{P\sigma P}$, being $\mathbf{P} = {\rm diag} (1,1,1,-1)$. This bound expresses  in the phase-space language the Peres-Horodecki PPT (Positive Partial Transpose) criterion for separability \cite{Peres:96, Horodecki:97, Simon:14, PhysRevA.72.032334}.

Therefore, the smallest symplectic eigenvalue encodes all the information needed to quantify the entanglement for arbitrary two-modes Gaussian states. For example, one can measure the entanglement through the violation of the PPT  criterion \cite{PhysRevLett.90.027901}. Quantitatively, this is given by the logarithmic negativity of a quantum state $\varrho$, which -- in the continuous variables formalism -- can be computed considering the following formula \cite{PhysRevA.65.032314, PhysRevA.72.032334}:
\begin{align}
\label{eq:log_neg}
E_{\mathcal{N}} ( \varrho) = {\rm max} \left [ 0, - \ln{\tilde{\nu}_{-}}\right ] .
\end{align}
Given the global state $\varrho$ and the two single-mode states $\varrho_{i} = {\rm Tr}_{j \ne i} \varrho$,  global $\mu \equiv {\rm Tr} \varrho^2$ and the local  $\mu_{1,2} \equiv {\rm Tr} \varrho_{1,2}^2$ purities can be used to characterise  entanglement in Gaussian systems. It has been shown that two different classes of extremal states can be identified: states of maximum negativity for fixed global and local purities (GMEMS) and states of minimum negativity for fixed global and local purities (GLEMS) \cite{PhysRevLett.92.087901}.

Moreover, the continuous variables approach provides a remarkable advantage:  the open quantum system dynamics can be remapped into a Fokker-Plank equation for the Wigner function of the system. This formal result enables us to carry out our study of the entropy production using a different approach based on phase-space methods, instead of resorting to the usual approach based on von Neumann entropy. The harmonic nature of the system we would like to consider makes our choice perfectly appropriate to our study and, as we will show in \Cref{sec:main,sec:Markovian_Limit}, well suited to systematically scrutinise inter-system correlations. Our analysis is thus based on the Wigner entropy production rate \cite{PhysRevLett.118.220601}, defined as
\begin{align}
\Pi(t) \equiv - \partial_t K(W(t) || W_{\rm eq}),
\end{align}
where $K(W|| W_{\rm eq})$ is the Wigner relative entropy between the Wigner function $W$ of the system and its expression for the equilibrium state $W_{\rm eq}$.

Furthermore, we are in the position of using the closed expressions for $\Phi(t)$ and $\Pi(t)$ coming from the theory of classical stochastic processes \cite{Brunelli, Landi2013}. In particular, it has been shown that the entropy production rate $\Pi(t)$ can be expressed in terms of the matrices $\mathbf{A},\mathbf{D},\boldsymbol{\sigma}$ as~\cite{Brunelli}
\begin{equation}
\label{eq:entropy_prod_rate}
\begin{aligned}
\Pi(t) &= \frac{1}{2} {\rm{Tr}} [ \boldsymbol{\sigma}^{-1} \mathbf{D}] + 2 {\rm{Tr}} [ \mathbf{A}^{{\rm irr}}]   \\ 
&+ 2 {\rm{Tr}} [ \left (\mathbf{A}^{{\rm irr}}\right )^{{\rm T}} \mathbf{D}^{-1} \mathbf{A}^{{\rm irr}} \boldsymbol{\sigma}] \ ,
\end{aligned}
\end{equation}
where $\mathbf{A}^{{\rm irr}}$ is the irreversible part of matrix $\mathbf{A}$, given by
$\mathbf{A}^{{\rm irr}} =  \left ( \mathbf{A} + \mathbf{E} \mathbf{A} \mathbf{E}^{{\rm T}}\right )/2$,
where $\mathbf{E} = {\rm diag}(1,-1,1-1)$ is the symplectic representation of time reversal operator.

\section{Quantum Brownian motion}
\label{sec:QBM}
We study the relation between the preparation of the initial state and the entropy production rate considering a rather general example: the quantum Brownian motion \cite{Breuer-Petruccione, Weiss1999}, also known as Caldeira-Leggett model \cite{CaldeiraLeggett1983a} . More specifically, we consider the case of a harmonic oscillator interacting with a bosonic reservoir made of independent harmonic oscillators. The study of such a paradigmatic system has been widely explored in both the Markovian \cite{Breuer-Petruccione, CaldeiraLeggett1983a} and non-Markovian \cite{PhysRevD.45.2843} regimes using the influence functional method: in this case, one can trace out the environmental degrees of freedom exactly. 
One can also solve the dynamics of this model using the open quantum systems formalism \cite{Breuer-Petruccione, Rivas2012}, where the Brownian particle represents the system, while we identify the bosonic reservoir with the environment. 
The usual approach relies on the following set of assumptions, which are collectively known as Born-Markov approximation~\cite{Breuer-Petruccione}:
\begin{enumerate}
\item The system is weakly coupled to the environment.
\item The initial system-environment state is factorised.
\item \label{Born-Markov3} It is possible to introduce a separation of the timescales governing the system dynamics and the decay of the environmental correlations.
\end{enumerate}
However, we aim to solve the dynamics in a more general scenario, without resorting to assumption \ref{Born-Markov3}. We are thus considering the case in which, although the system-environment coupling is weak, non-Markovian effects may still be relevant. Under such conditions, one can derive a time-local master equation for the reduced dynamics of the system \cite{PhysRevA.67.042108, Int1}.

\begin{figure}
\centering
\includegraphics[width=0.83\linewidth]{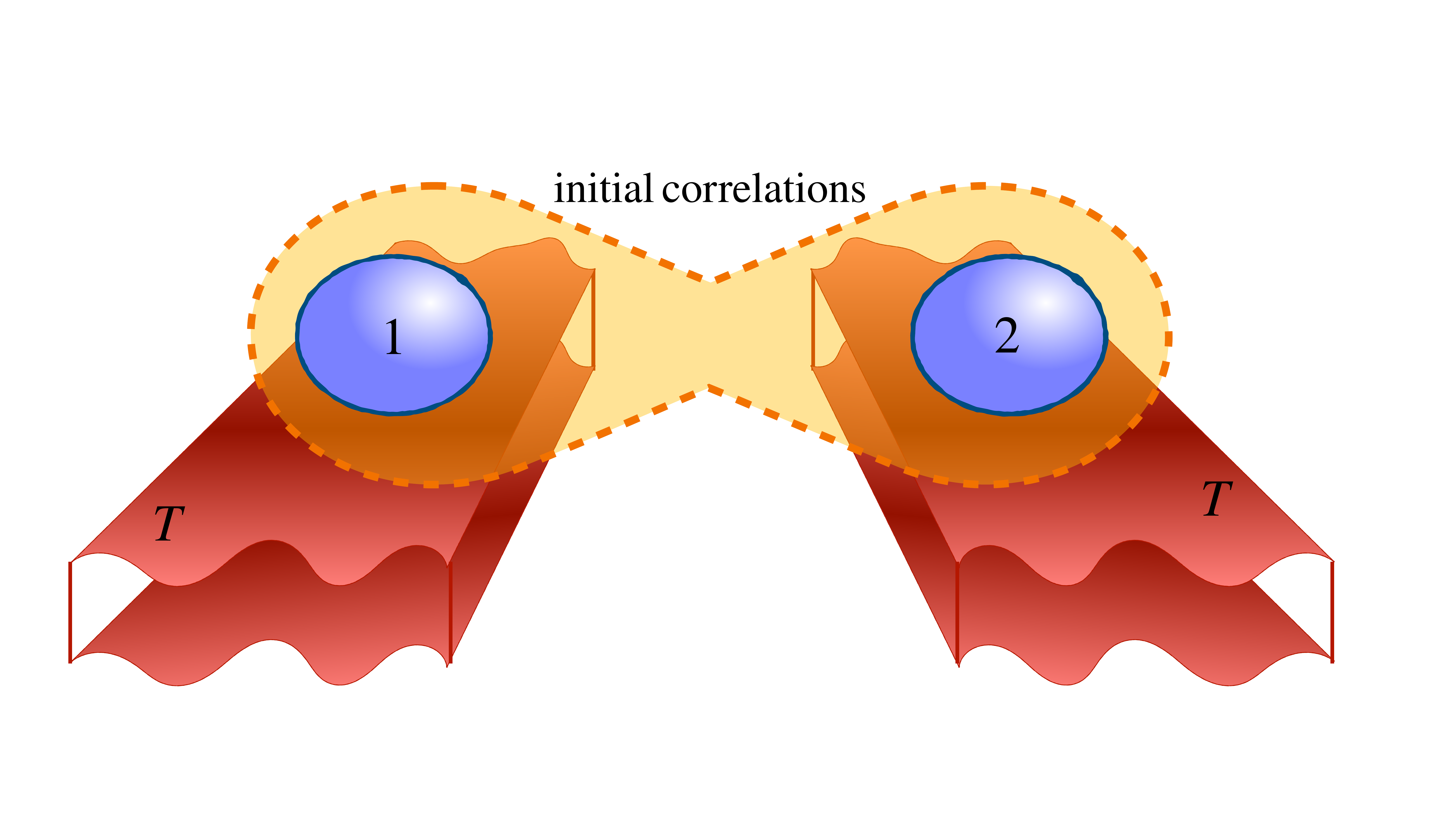}
\caption{\small{System of two uncoupled quantum harmonic oscillators interacting with their local reservoirs. The latter are characterised by the same temperature $T$ and the same spectral properties. The two parties of the systems are initially correlated and we study their dynamics under the secular approximation so that non-Markovian effects are present.}}
\label{fig:system}
\end{figure}

More specifically, we consider a system consisting of two quantum harmonic oscillators, each of them interacting with its own local reservoir (see \Cref{fig:system}). Each of the two reservoirs is modelled as a system of system of $N$ non-interacting bosonic modes. 
In order to understand the dependence of the entropy production upon the initial correlations, we choose the simplest case in which the two oscillators are identical, i.e., characterised by the same bare frequency $\omega_0$ and the same temperature $T$, and they are uncoupled, so that only the initial preparation of the global state may entangle them. The Hamiltonian of the global system thus reads as (we consider units such that $\hbar = 1$ throughout the paper)
\begin{align}
\label{eq:Hamiltonian}
H & = \sum_{j=1,2} \omega_0 \, a_{j}^{\dagger} a_j + \sum_{j=1,2} \sum_k \omega_{jk} \, b_{jk}^{\dagger} b_{jk} \nonumber \\
& + \alpha \sum_{j=1,2} \sum_k \left ( \frac{a_j + a_j^{\dagger}}{\sqrt{2}}\right ) \left ( g_{jk}^{*} b_{jk} + g_{jk} b_{jk}^{\dagger} \right ),
\end{align}
where $a_j^{\dagger}$ ($a_j$) and $b_{jk}^{\dagger}$ ($b_{jk}$) are the system and reservoirs creation (annihilation) operators, respectively, while $\omega_{1k}$ and $\omega_{2k}$ are the frequencies of the reservoirs modes. The dimensionless constant $\alpha$ represents the coupling strength between each of the two subsystems and the their local bath, while the constants $g_{jk}$ quantify the coupling between the $j-$th oscillator ($j=1,2$) and the $k-$th mode of its respective reservoir. These quantities therefore appear in the definition of the spectral density (SD)
\begin{align}
\label{eq:SD-def}
J_{j} (\omega) = \sum_{k} |g_{jk}|^2 \, \delta (\omega - \omega_{jk}) \ .
\end{align}
In what follows, we will the consider the case of symmetric reservoirs, i.e., $J_1(\omega) = J_2(\omega) \equiv J(\omega)$. 

We would also like to work in the secular approximation by averaging over the fast oscillating terms after tracing out the environment: unlike the rotating-wave approximation, in this limit not all non-Markovian effects are washed out \cite{PhysRevA.67.042108}.

Under these assumptions, the dynamics of this system is governed by a time-local master equation, that in the interaction picture reads as
\begin{equation}
\label{eq:ME_sec}
\begin{aligned}
\dot{\rho} (t) = &-  \frac{\Delta(t) + \gamma(t)}{2} \sum_{j=1,2} \left ( \{\adag_j a_j, \rho\} - 2 a_j \rho \adag_j \right ) \\
& - \frac{\Delta(t) - \gamma(t)}{2} \sum_{j=1,2}  \left (\{a_j \adag_j, \rho \}- 2 \adag_j \rho a_j \right ),
\end{aligned}
\end{equation}
where $\rho$ is the reduced density matrix of the global system, while the time dependent coefficients $\Delta(t)$ and $\gamma(t)$ account for diffusion and dissipation, respectively. 
The coefficients in \Cref{eq:ME_sec} have a well-defined physical meaning: $(\Delta(t) + \gamma(t))/2$ is the rate associated with the incoherent loss of excitations from the system, while $(\Delta(t) - \gamma(t))/2$ is the rate of incoherent pumping.

The coefficients $\Delta(t)$ and $\gamma(t)$ are ultimately related to the spectral density $J(\omega)$ as
\begin{align}
\label{eq:QBM_delta}
\Delta(t) \equiv \frac{1}{2} \int_{0}^{t}  \kappa (\tau) \cos{(\omega_0 \tau}) \, d \tau,
\end{align}
\begin{align}
\label{eq:QBM_gamma}
\gamma (t) \equiv \frac{1}{2} \int_{0}^{t}  \mu (\tau)  \sin{(\omega_0 \tau})  \, d \tau,
\end{align}
where $\kappa(\tau)$ and $\mu(\tau)$ are the noise and dissipation kernels, respectively, which -- assuming reservoirs in thermal equilibrium -- are given by  
\begin{equation}
\label{eq:kappamu}
\left[
\begin{matrix}
\kappa(\tau)\\
\mu(\tau)
\end{matrix}\right]=
2 \alpha^2 \int\limits_{0}^{\mathcal{1}} J (\omega) \left[\begin{matrix}
\cos{(\omega \tau)} \coth{\left(\frac{\beta}{2} \omega \right)}\\
\sin(\omega\tau)\end{matrix}\right] d \, \omega,
\end{equation}
where $\beta = (k_B T)^{-1}$ is the inverse temperature and $k_B$ the Boltzmann constant.

Moreover, it can be shown that the dynamics of a harmonic system that is linearly coupled to an environment can be described in terms of a differential equation in the Lyapunov form given by \Cref{eq:Lyapunov} \cite{Serafini:17}. We can indeed notice that in \Cref{eq:Hamiltonian} the interaction between each harmonic oscillator and the local reservoir is expressed by a Hamiltonian that is bilinear (i.e., quadratic) in the system and reservoir creation and annihilation operators. Hamiltonians of this form lead to a master equation as in \Cref{eq:ME_sec}, where the dissipators  are quadratic in the system creation and annihilation operators $\adag_j, a_j$. Under these conditions, one can recast the dynamical equations in the Lyapunov form in \Cref{eq:Lyapunov}  \cite{Ferraro:05}, where the matrices $\mathbf{A}$ and $\mathbf{D}$ are time-dependent, due to non-Markovianity. Indeed, we get $\mathbf{A} = -\gamma(t) \mathbbm{1}_4$ and $\mathbf{D}= 2 \Delta(t)  \mathbbm{1}_4$ (here $ \mathbbm{1}_4$ is the $4 \times 4$ identity matrix).

The resulting Lyapunov equation can be analytically solved, giving the following closed expression for the CM at a time $t$:
\begin{align}
\label{eq:sigma_t}
\boldsymbol{\sigma}(t) = \boldsymbol{\sigma}(0) e^{-\Gamma(t)} + 2 \Delta_{\Gamma}(t) \mathbbm{1}_4,
\end{align}
with
\begin{equation}
\begin{aligned}
\label{eq:del_gamma}
\Gamma(t) \equiv 2 \int_{0}^{t}  d \tau \,  \gamma(\tau)\,\,
\text{and}\,\,
\Delta_{\Gamma} (t) \equiv e^{-\Gamma(t)} \int_{0}^{t} d \tau \, \Delta(\tau) e^{\Gamma(\tau)}.
\end{aligned}
\end{equation}
Moreover, a straightforward calculation allows us to determine the steady state of our two-mode system. By imposing $\dot{\boldsymbol{\sigma}} \equiv 0$ in  \Cref{eq:Lyapunov}, one obtains that the system relaxes towards a diagonal state with associated CM ${\boldsymbol{\sigma}}_{\mathcal{1}} \equiv \Delta(\infty)/\gamma(\infty) \mathbbm{1}_4$. By plugging ${\boldsymbol{\sigma}}_{\mathcal{1}}$ in \Cref{eq:entropy_prod_rate}, we find $\Pi_\infty\equiv\lim_{t\rightarrow \infty}\Pi(t)=0$, showing a vanishing entropy production at the steady state. This instance can also be justified by noticing that, as $t \to \mathcal{1}$, we approach the Markovian limit. Therefore, the Brownian particles, exclusively driven by the interaction with their local thermal baths, will be relaxing toward the canonical Gibbs state with a vanishing associated entropy production rate \cite{Breuer-Petruccione, Spohn1978, Landi:19}.

\subsection{Choice of the spectral density} 
In order to obtain a closed expression for the time-dependent rates $\Delta(t)$ and $\gamma(t)$, one has to assume a specific form for the spectral density $J(\omega)$, which -- to generate an irreversible dynamics -- is assumed to be a continuous function of the frequency $\omega$. In quite a general way, we can express the SD as
\begin{align}
J(\omega) = \eta \ \omega_c^{1-\epsilon} \ \omega^\epsilon \; f(\omega, \omega_c),
\end{align}
where $\epsilon>0$ is known as the Ohmicity parameter and $\eta > 0$. Depending on the value of $\epsilon$, the SD is said to be Ohmic ($\epsilon=1$), super-Ohmic ($\epsilon>1)$, or sub-Ohmic ($\epsilon<1$). The function $f(\omega, \omega_c)$ represents the SD cut-off and $\omega_c$ is the cut-off frequency. Such  function is introduced so that  $J(\omega)$ vanishes for $\omega \to 0$ and $\omega \to \mathcal{1}$. We focus on two different functional forms for $f(\omega, \omega_c)$, namely, the Lorentz-Drude cut-off $f(\omega, \omega_c) \equiv \omega_c^2 / (\omega_c^2 + \omega^2)$ and the exponential cut-off $f(\omega, \omega_c) \equiv e^{-\omega/\omega_c}$.
In particular, we choose an Ohmic SD with a Lorentz-Drude cut-off
\begin{align}
\label{eq:SD_Ohm_LD}
J(\omega) = \frac{2 \omega}{\pi} \frac{\omega_c^2}{\omega_c^2 + \omega^2},
\end{align}
where $\eta \equiv 2 / \pi$. Note that this choice is mathematically convenient, but is inconsistent from a physical point of view, as it implies instantaneous dissipation, as acknowledged in Refs.~\cite{PhysRevD.45.2843, PazZurek2001}.
We also consider the following SDs
\begin{align}
\label{eq:SD_exp}
J(\omega) =  \omega_c^{1-\epsilon} \ \omega^\epsilon \; e^{-\omega/\omega_c},
\end{align}
with $\epsilon=1, \, 3, \,1/2$ and $\eta \equiv 1$, as the coupling strength is already contained in the constant $\alpha$. In all these cases, the time-dependent coefficients $\Delta(t)$ and $\gamma(t)$ can be evaluated analytically~\cite{PhysRevA.80.062324}.

\section{Initial correlations and entropy production}
\label{sec:main}
We can now use our system to claim that initial correlations shared by the non-interacting oscillators do play a role in the entropy production rate. We do this by employing a parametrisation that covers different initial preparations~\cite{PhysRevA.72.032334}.  The entries of the matrix given by \Cref{eq:sigma_sf} can be expressed as follows
\begin{equation}
\label{eq:sigma_a_b}
a = s +d, \qquad b = s-d
\end{equation}
and
\begin{equation}
\begin{aligned}
\label{eq:sigma_c}
c_{\pm} = \frac{ \sqrt{\left (4d^2 + f \right )^2{-}4g^2} \pm  \sqrt{\left (4s^2 + f  \right )^2{-}4g^2} }{4 \sqrt{s^2 -d^2}},
\end{aligned}
\end{equation}
with $f = (g^2 +1)(\lambda -1) /2- (2d^2+g)(\lambda +1)$. This allows us to parametrise the CM using four parameters: $s, d, g,  \lambda$. The local purities are controlled by the parameters $s$ and $d$ as $\mu_1 = (s+d)^{-1}$ and $\mu_2 = (s-d)^{-1}$, while the global purity is $\mu=1/g$. Furthermore, in order to ensure  legitimacy of a CM, the following constraints should be fulfilled
\begin{align}
\label{eq:constraints}
s \ge 1, \quad | d | \le s -1, \quad g \ge 2| d | + 1.
\end{align}
Once the three aforementioned purities are given, the remaining degree of freedom required to determine the negativities is controlled by the parameter $\lambda$, which encompasses all the possible entangled two-modes Gaussian states. The two classes of extremal states are obtained upon suitable choice of  $\lambda$. For $\lambda=-1$ ($\lambda = +1$) we recover the GLEMS (GMEMS) mentioned in~\Cref{sec:Gauss}. 

\begin{figure}
\centering
\includegraphics[width=0.83\linewidth]{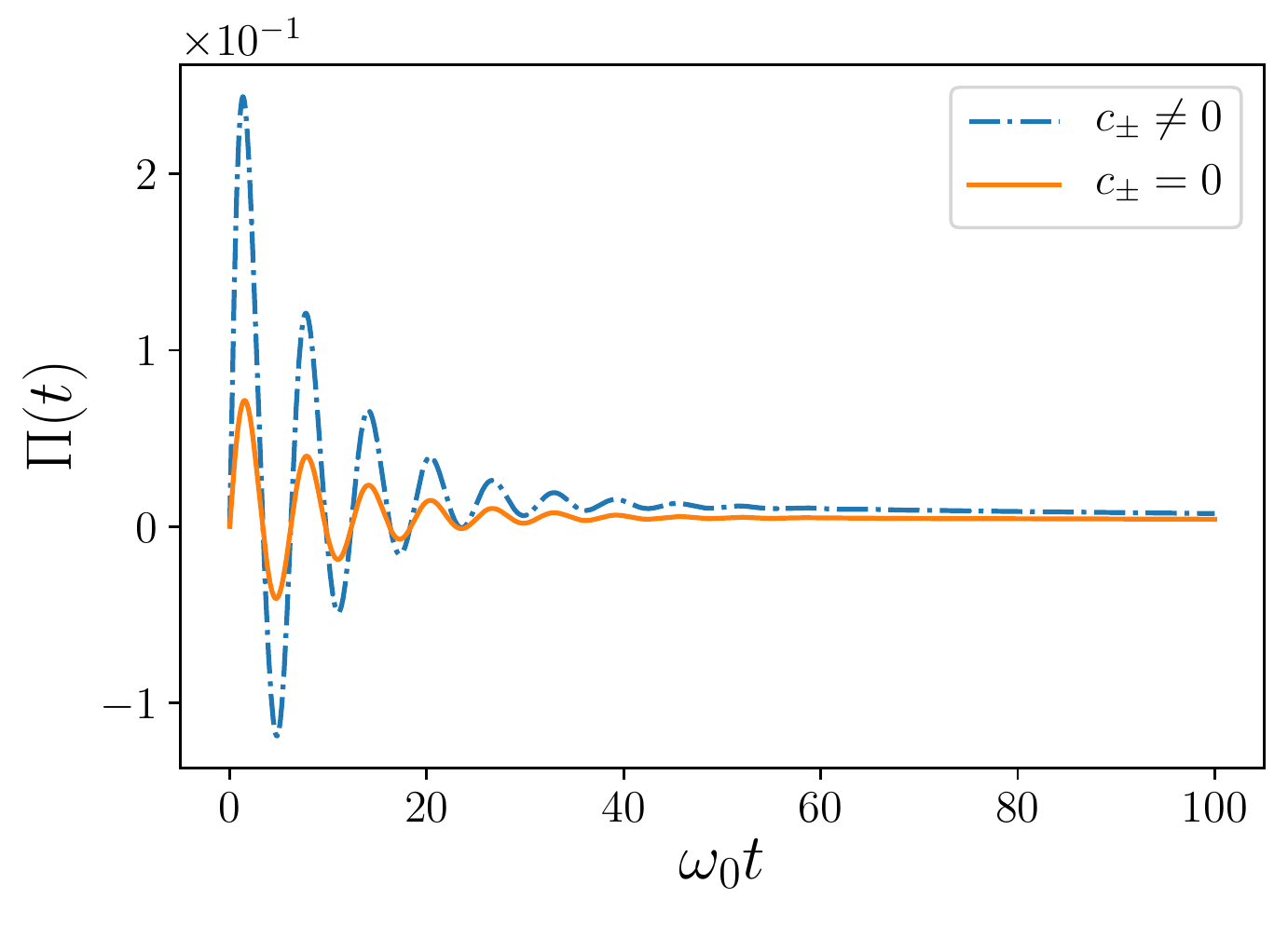}
\caption{\small{Entropy production rate in a system of two non-interacting oscillators undergoing the non-Markovian dynamics described in~\Cref{sec:QBM}. We compare the behaviour of the entropy production rate resulting from a process where the system is initialised in a state with no initial correlations (solid line) to what is obtained starting from a correlated state (dashed-dotted line). The latter case refers to the preparation of a system in a pure ($g=1$), symmetric ($d=0$) squeezed state ($\lambda = 1$). The former situation, instead, corresponds to taking the tensor product of the local states. In this plot we have taken $s=2$ and an Ohmic SD with Lorentz-Drude cut-off. The system parameters are  $\alpha = 0.1 \omega_0$, $\omega_c = 0.1 \omega_0$, $\beta = 0.1 \omega_0^{-1}$.}}
\label{fig1}
\end{figure}

To show a preview of our results, we start with a concrete case shown in \Cref{fig1}. We prepare the system in a pure ($g=1$) symmetric ($d=0$) state, and investigate the effects of initial correlations on $\Pi(t)$ by comparing the value taken by this quantity for such an initial preparation with what is obtained by considering the covariance matrix associated with the tensor product of the local states of the oscillators, i.e., by forcefully removing the correlations among them. Non-Markovian effects are clearly visible in the oscillations of the entropy production and lead to negative values of $\Pi(t)$ in the first part of the evolution. This is in stark contrast with the Markovian case, which entails non-negativity of the entropy production rate. Crucially, we see that, for a fixed initial value of the local energies, the presence of initial correlations enhances the amount of entropy produced at later times, increasing the amplitude of its oscillations. We also stress that both curves eventually settle to zero (on a longer timescale than shown in \Cref{fig1}) as argued in Sec.~\ref{sec:QBM}.

\begin{figure}
\centering
\includegraphics[width=0.83\linewidth]{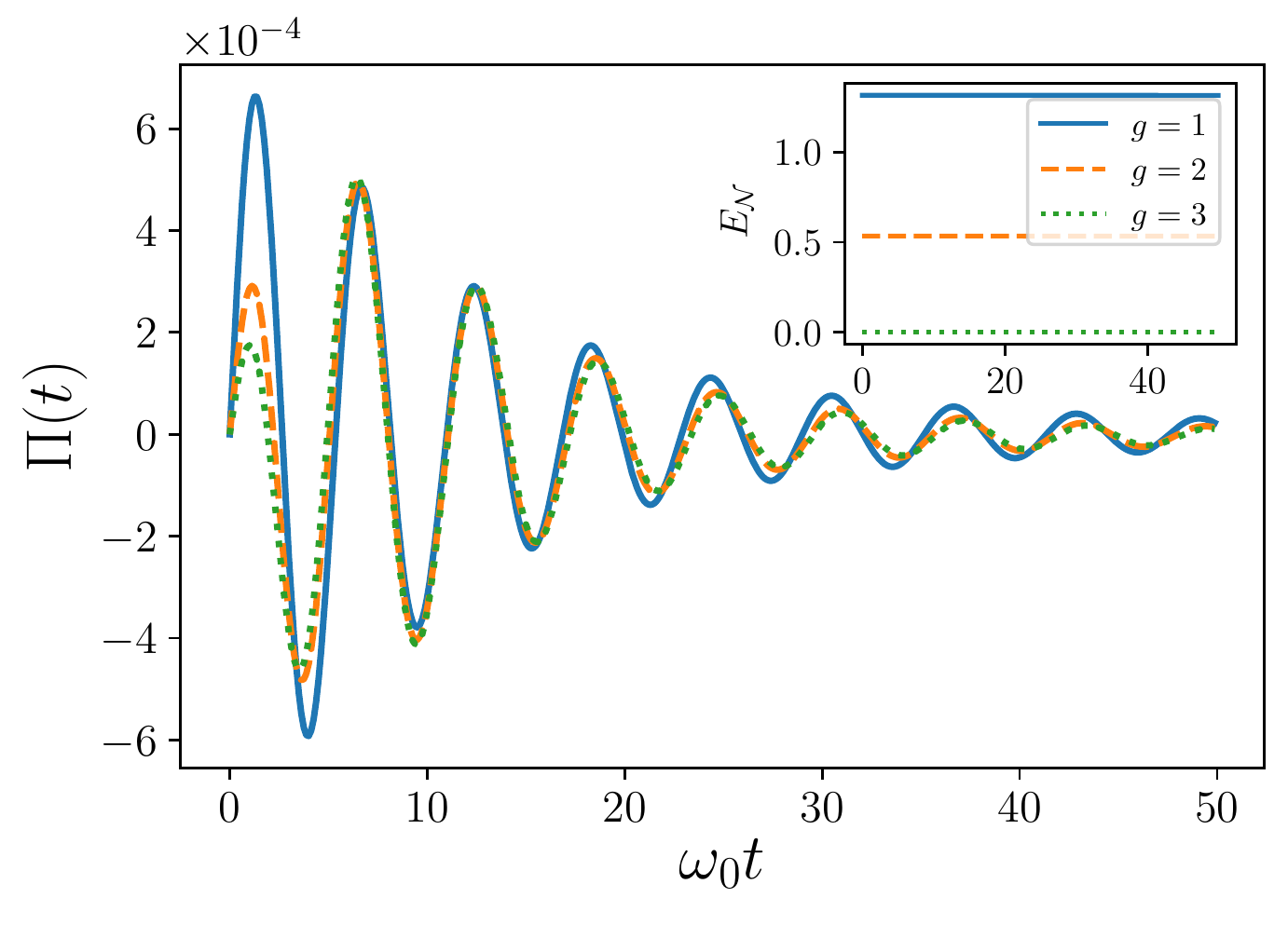}
\caption{\small{Entropy production rates corresponding to independent zero-temperature reservoirs. We consider preparations of the initial global state corresponding to different values of parameter $g$ (related to the global purity of the state), while fixing $s=2$, $d=0$, and $\lambda = 1$. In the inset we plot the logarithmic negativity for the same choice of parameters: entanglement persists over time up to the reach of a steady state of the dynamics. We have taken an Ohmic SD with an exponential cut-off with  $\alpha = 0.1 \omega_0$, $\omega_c = 0.1 \omega_0$.}}
\label{fig_vacuum}
\end{figure}

We now move to a more systematic investigation of $\Pi(t)$ and its dependence on the specific choice of  $s, d, g,  \lambda$.
In order to separate the contributions, we first study the behaviour of $\Pi(t)$ when we vary one of those parameters, while all the others are fixed. We can first rule out the contribution of thermal noise by considering the case in which the reservoirs are in their vacuum state. Such zero-temperature limit can be problematic, as some approaches to the quantification of entropy production fail to apply in this limit~\cite{PhysRevLett.118.220601}. In contrast, phase-space methods based on the R\'{e}nyi-$2$-Wigner entropy allow to treat such a limit without pathological behaviours associated with such {\it zero-temperature catastrophe}~\cite{PhysRevLett.118.220601}. This formal consistency is preserved also in the case of a system whose dynamics is described by \Cref{eq:ME_sec}, as shown in \Cref{fig_vacuum}.
We take $T=0$ and choose an Ohmic SD with an exponential cut-off -- given by \Cref{eq:SD_exp} -- with $\epsilon=1$. 
The map describing the dynamics converges to a stationary state characterised by a vanishing $\Pi(t)$,  although the oscillations are damped to zero more slowly, as non-Markovian effects are more persistent in the presence of zero-temperature reservoirs. Furthermore, we notice that the differences between different initial states are most pronounced in correspondence of the first peak: this suggests that the maximum value for the entropy production can be reasonably chosen as an apt figure of merit to distinguish the differences due to state preparation. 
Supported by this evidence, we adopt the value of the first maximum of $\Pi(t)$ as an indicator of the irreversibility generated in the relaxation dynamics by different initial preparations.

In the inset of \Cref{fig_vacuum} we show the logarithmic negativity given by \Cref{eq:log_neg}. The interaction with zero-temperature reservoirs does not cause detrimental effects to entanglement, as the latter is preserved over time \cite{PhysRevLett.100.220401,PhysRevA.75.062119, PhysRevA.80.062324}.

\begin{figure}
\centering
\includegraphics[width=0.83\linewidth]{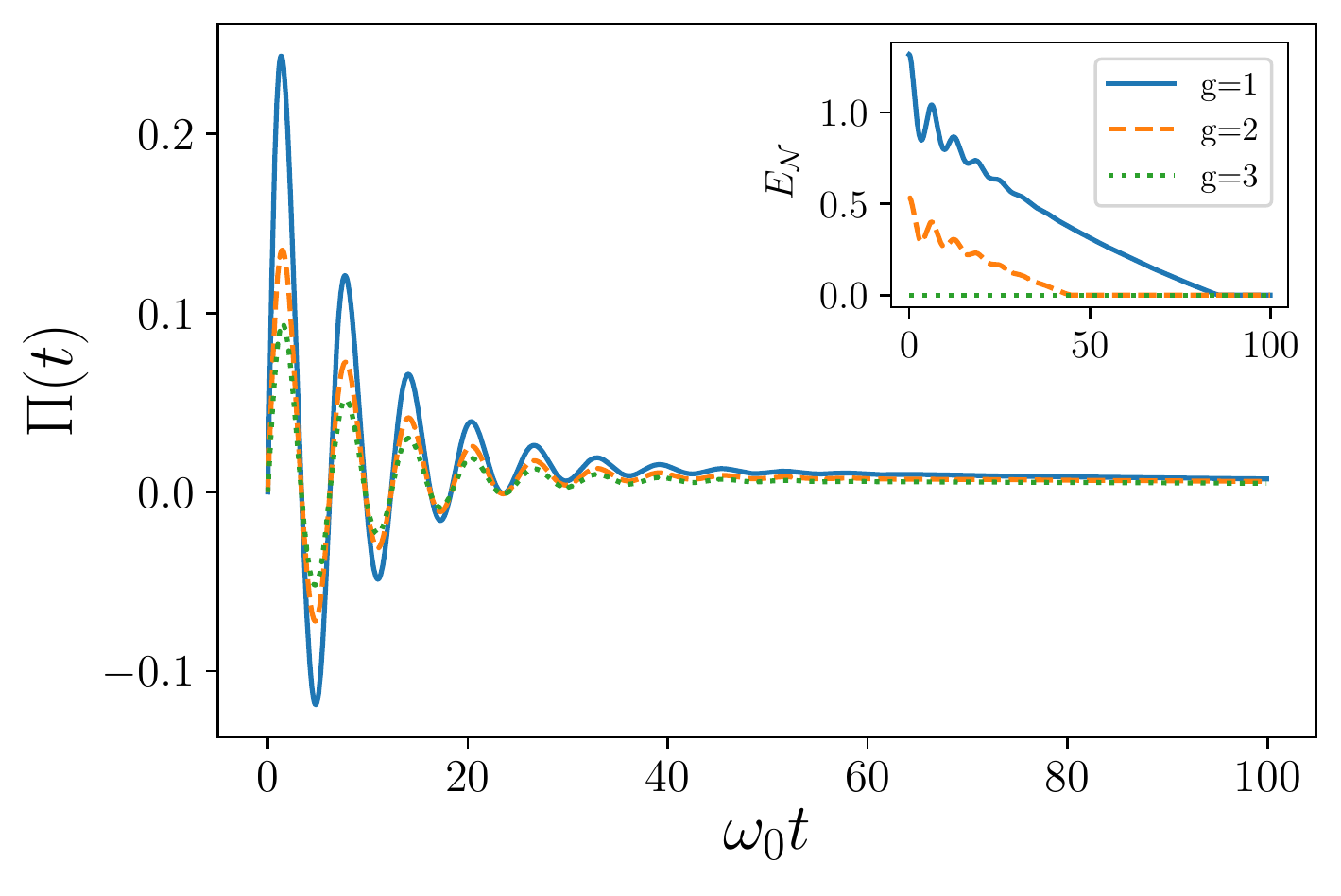}
\caption{\small{Entropy production rates corresponding to different preparations of the initial global state. We consider different values of parameter $g$, while taking $s=2$, $d=0$, and $\lambda = 1$. In the inset we plot the logarithmic negativity for the same choice of the parameters. The dynamics of the system has been simulated using an Ohmic SD with a Lorentz-Drude cut-off. The system parameters are  $\alpha = 0.1 \omega_0$, $\omega_c = 0.1 \omega_0$, $\beta = 0.1 \omega_0^{-1}$.}}
\label{fig2}
\end{figure}

We also address the case of finite-temperature reservoirs and an Ohmic SD with Lorentz-Drude cut-off given by \Cref{eq:SD_Ohm_LD}~\footnote{the analysis can easily be extended to the exponential cut-off in \Cref{eq:SD_exp} for the Ohmic ($\epsilon=1$), super-Ohmic ($\epsilon=3$) and sub-Ohmic ($\epsilon=1/2$) case}. 
We thus fix $s, d, \lambda$ and  let $g$ vary to explore the role played by the global purity. \Cref{fig2} shows that, by increasing $g$ -- i.e., by reducing the purity of the global state -- $\Pi(t)$ decreases: an initial state with larger purity lies far from an equilibrium state at the given temperature of the environment and is associated with a larger degree of initial entanglement [cf. inset of \Cref{fig2} and the analysis reported in \Cref{sec:entanglement}], which translates in a larger entropy production rate. Furthermore, our particular choice of the physical parameters leads to the observation of ``entanglement sudden death''~\cite{PhysRevLett.100.220401, PhysRevA.80.062324}: an initial state with non-null logarithmic negativity completely disentangles in a finite time due to interaction with environment, the disentangling time being shortened by a growing $g$ [cf. inset of \Cref{fig2}].

\begin{figure}[b]
\centering
\includegraphics[width=0.83\linewidth]{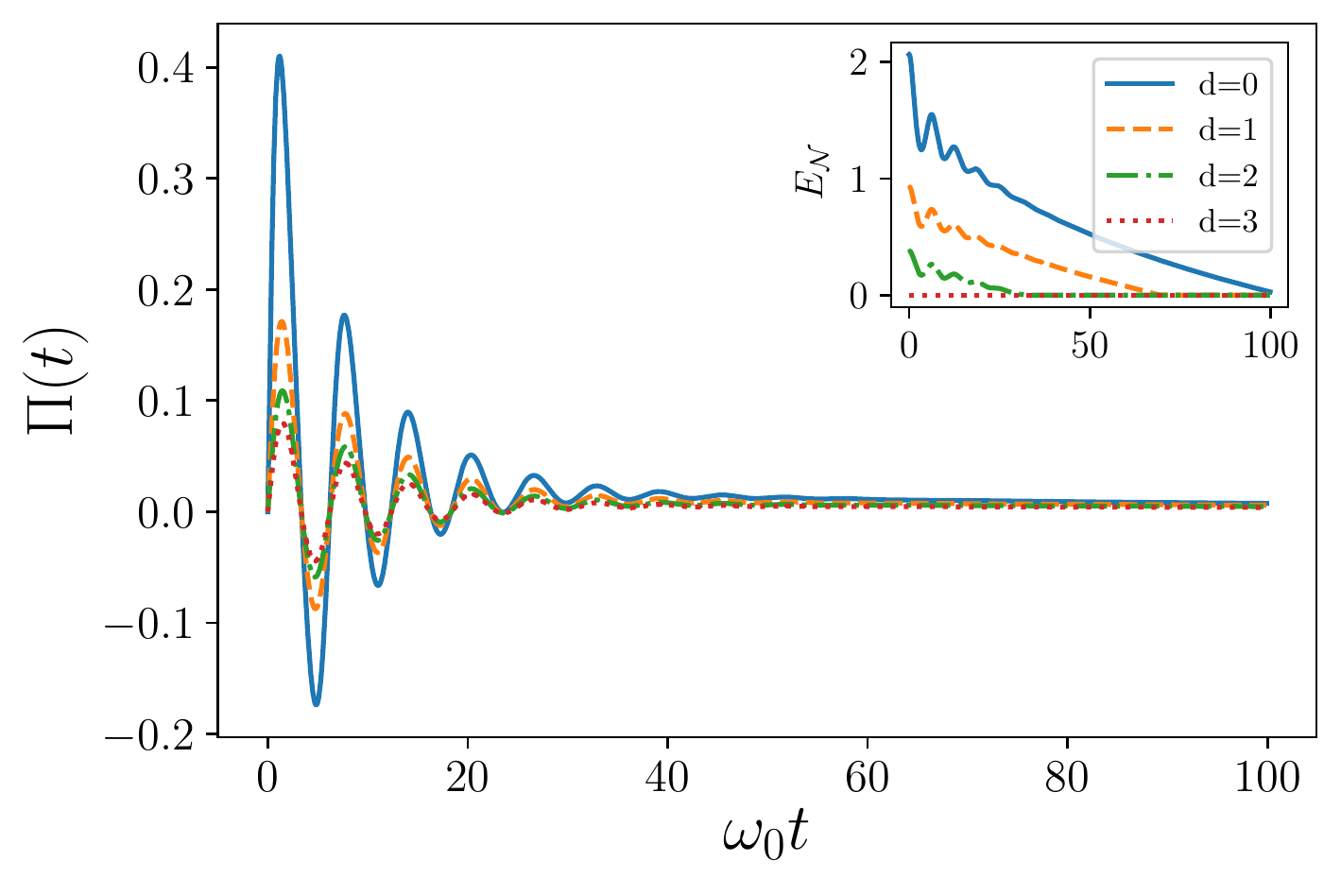}
\caption{\small{Entropy production rates corresponding to different values of $d$ and $g=2d+1$ in the parametrisation of the initial state (we have taken $s=4$ and $\lambda = 1$). In the inset, the behaviour of the logarithmic negativity is shown. In this figure, we take an Ohmic SD with a Lorentz-Drude cut-off and $\alpha = 0.1 \omega_0$, $\omega_c = 0.1 \omega_0$, $\beta = 0.1 \omega_0^{-1}$.}}
\label{fig3}
\end{figure}

Similarly, we can bias the local properties of the oscillators by varying $d$ and, in turn, $g= 2d +1$, while keeping $s, \lambda$ fixed: in \Cref{fig3} we can observe that, when the global energy is fixed, the asymmetry in the local energies -- and purities $\mu_1$ and $\mu_2$ -- reduces the entropy production rate. In the inset we show that, by increasing the asymmetry between the two modes, the entanglement takes less time to die out. These results are consistent with the trends observed in \Cref{fig2}.
Indeed, a bias in the local energies would make the reduced state of one of the two oscillators more mixed, and thus less prone to preserve the entanglement that is initially set in the joint harmonic state. Such imbalance would give different weights to the two local dissipation processes, thus establishing an effective preferred local channel for dissipation. In turn, this would result in a lesser weight to the contribution given by correlations.

\begin{figure}[t]
\centering
\includegraphics[width=0.83\linewidth]{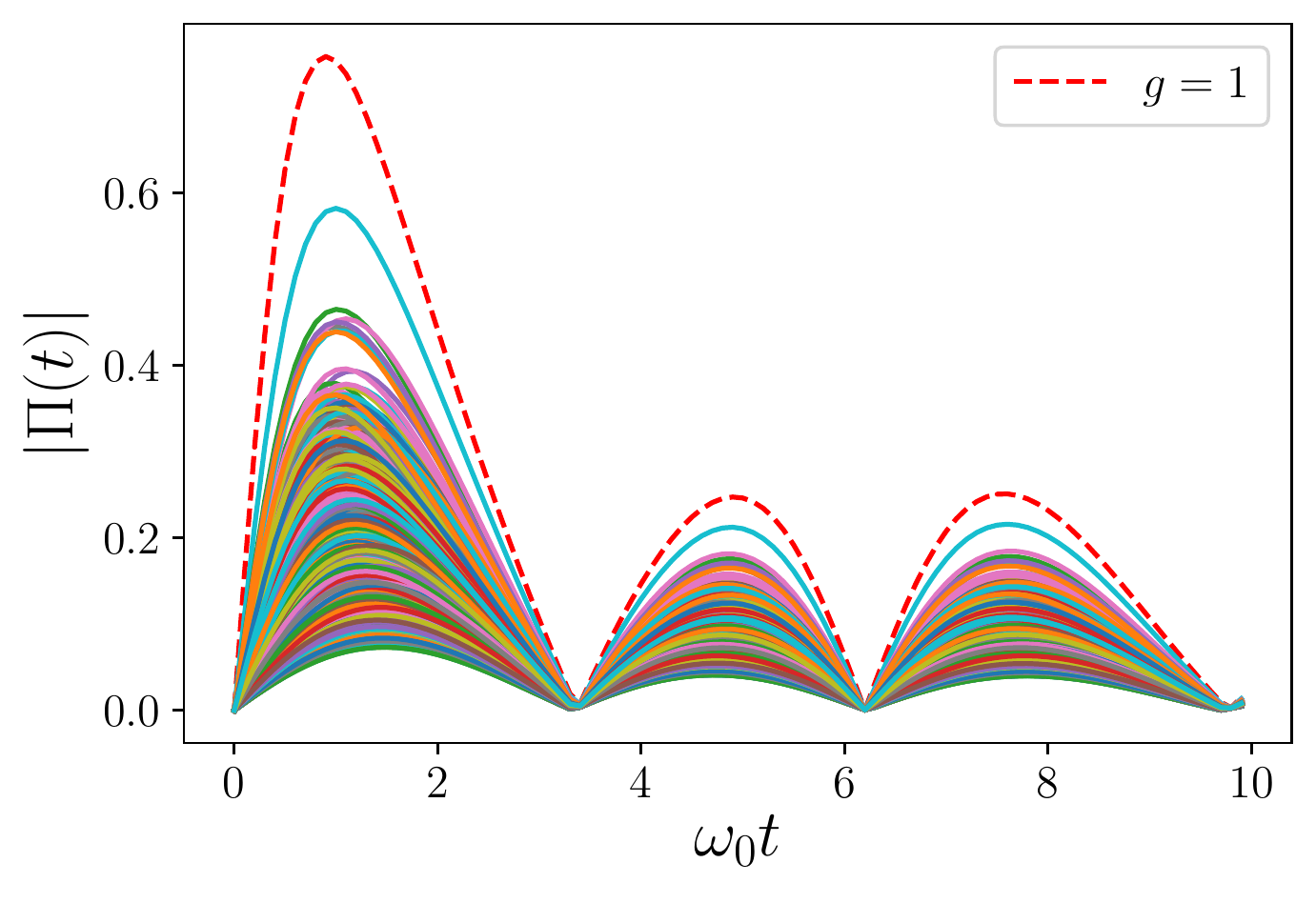}
\caption{\small{Entropy production rates $\Pi(t)$ (absolute value)  as a function of time. The initial CM is parametrised by fixing $s$ ($s=10$ in the figure) and randomly choosing $d, g, \lambda$ such that they are uniformly distributed in the intervals $[0,s-1]$, $[2d+1,d+10]$ and $[-1,1]$ respectively. The figure reports $N_{R} = 1000$ different realisations of the initial state. The dashed line corresponds to the globally pure state ($g=1$) with $d=0$, $\lambda=1$. All the plots are obtained considering an Ohmic SD with a Lorentz-Drude cut-off and $\alpha = 0.1 \omega_0$, $\omega_c = 0.1 \omega_0$, $\beta = 0.1 \omega_0^{-1}$.}}
\label{fig4}
\end{figure}

We conclude our analysis in this Section by exploring the parameter space in a more systematic way by fixing the global energy $s$ and randomly choosing the three parameters left, provided that the constraints in \Cref{eq:constraints} are fulfilled. We see from \Cref{fig4} that the curve for $\Pi(t)$ comprising all the others is the one corresponding to unit global purity, i.e., $g=1$, and $d=0$, $\lambda=1$ (dashed line). The globally pure state is indeed the furthest possible from a diagonal one: the rate at which  entropy production varies is increased in order to reach the final diagonal state ${\boldsymbol{\sigma}}_{\mathcal{1}}$. 

\subsection{Dependence on the initial entanglement}
\label{sec:entanglement}

We now compare the trends corresponding to different choices of the parameters characterising the initial state. As non-Markovian effects are reflected in oscillating behaviour of the entropy production, we can contrast cases corresponding to different initial preparations by looking at the maximum and the minimum values $\Pi_{\rm max}$ and $\Pi_{\rm min}$ that the entropy production rate assumes for each choice of the parameters. Taking into account the evidence previously gathered, in the simulations reported in this Section we fix the minimum value for $g$, i.e., $g=2d+1$, and $\lambda = +1$ as significant for the points that we want to put forward. In fact, with such choices we are able to parametrise the initial state with a minimum number of variables, while retaining the significant features that we aim at stressing. We can further assume, without loss of generality, $d \ge 0$: this is simply equivalent to assuming that the first oscillator is initially prepared in a state with a larger degree of mixedness than the second one, i.e., $\mu_1 \le \mu_2$. In this case, we can express $d$ in terms of the smallest symplectic eigenvalue of the partially transposed CM $\tilde{\nu}_{-}$. Therefore, taking into account the constraints given by \Cref{eq:constraints}, one has that $d = -\frac{1}{2} ( \tilde{\nu}_{-}^2 - 2s \tilde{\nu}_{-} +1)$.
We already mentioned in \Cref{sec:QBM} that we are able to derive a closed expression for the CM at any time $t$, given by \Cref{eq:sigma_t}. We can further notice that the positive and negative peaks in the entropy production rate are attained at short times. We can thus perform a Taylor expansion of $\Delta(t)$ in \Cref{eq:del_gamma} to obtain
\begin{equation}
\begin{aligned}
\Delta_{\Gamma} (t) =   
[1- \Gamma(t)]  \int_{0}^{t} d \tau \Delta(\tau) + \int_{0}^{t} d \tau \; \Gamma(\tau) \Delta(\tau)  + \mathcal{O} (\alpha^4) \ .
\end{aligned}
\end{equation}
As $\Delta(t) \propto \alpha^2$ and $\Gamma(t) \propto \alpha^2$, we can retain only the first term consistently with the weak coupling approximation we are resorting to. Therefore, we can recast \Cref{eq:sigma_t} in a form that is more suitable for numerical evaluations, namely
\begin{align}
\label{eq:sigma_t_wc}
\boldsymbol{\sigma}(t) = \left [ 1 - \Gamma(t) \right ]  \boldsymbol{\sigma}(0) + \left [ 2 \int_{0}^{t} d \tau \Delta(\tau) \right ] \mathbbm{1}_4 \ .
\end{align}
By substituting \Cref{eq:sigma_t_wc} into \Cref{eq:entropy_prod_rate}, we get the analytic expression for the entropy production rate, given in \Cref{app:a} for the sake of completeness but whose explicit form is not crucial for our analysis here. 

\begin{figure}
\centering
\includegraphics[width=0.83\linewidth]{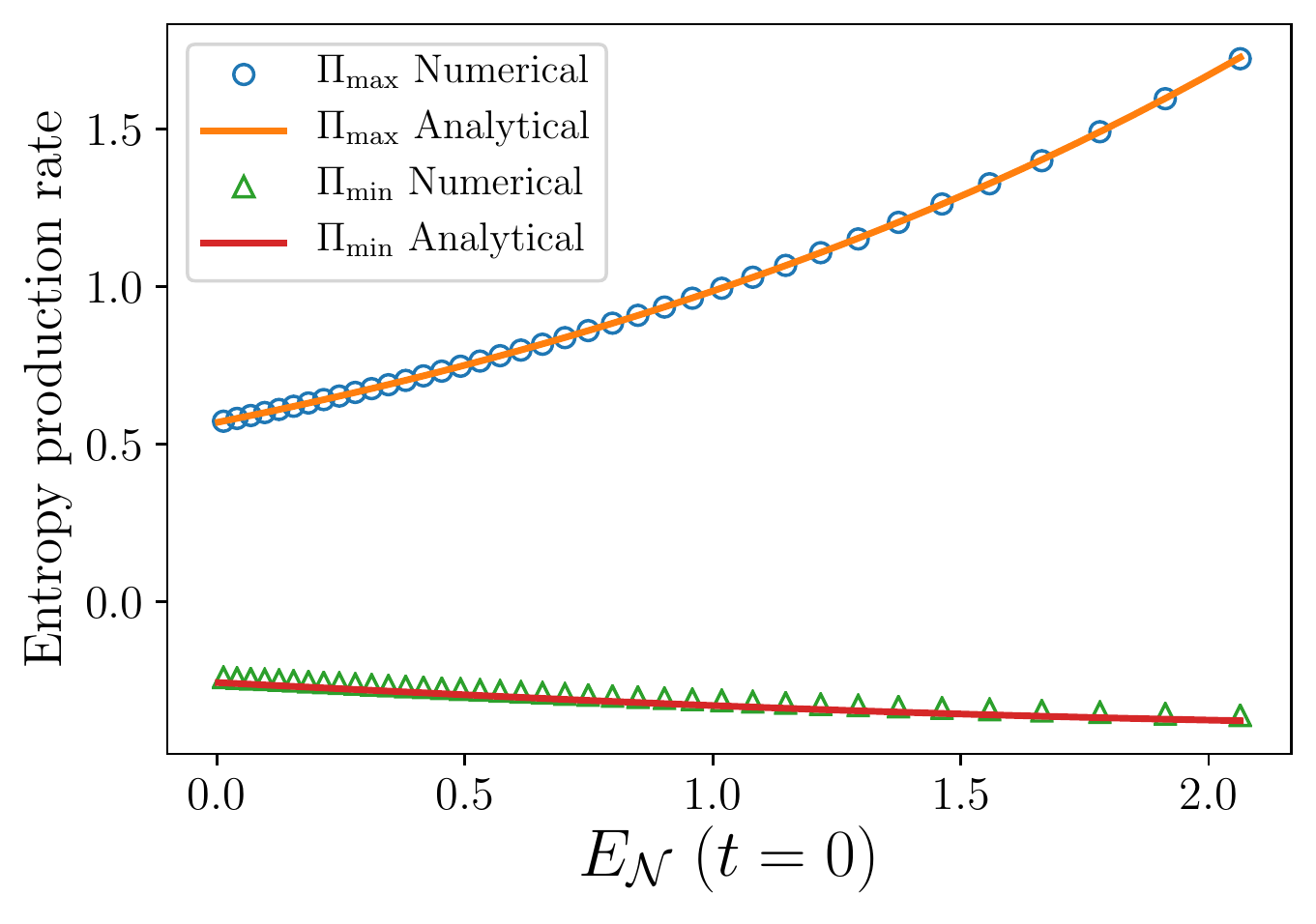}
\caption{\small{Maximum and minimum of the entropy production rate $\Pi_{\rm max}$ and $\Pi_{\rm min}$  as functions of the  entanglement negativity at $t=0$. We take $s=4$, $g=2d +1$, $\lambda = 1$, while $0 \le d \le 3$. We compare the numerical results (triangles and circles) to the analytical solution in \Cref{eq:entropy_prod_rate_general} (solid line). We have used an Ohmic SD with a Lorentz-Drude cut-off and $\alpha = 0.1 \omega_0$, $\omega_c = 0.1 \omega_0$, $\beta = 0.01 \omega_0^{-1}$.}}
\label{fig5}
\end{figure}

In this way, all the information about the initial state is encoded in the value of $\tilde{\nu}_{-}$ while $s$ is fixed. Note that this expression holds for any SD: once we choose the latter, we can determine the time-dependent coefficients $\Delta(t)$ and $\gamma(t)$ and thus the entropy production rate $\Pi(t)$. We can then compute the maximum of the entropy production rate and study the behaviour of $\Pi_{\rm max}$ and $\Pi_{\rm min}$ as functions of the entanglement negativity $E_{\mathcal{N}}$  at $t=0$. In \Cref{fig5} we compare numerical results to the curve obtained by considering the analytical solution discussed above and reported in \Cref{eq:entropy_prod_rate_general}. Remarkably, we observe a monotonic behaviour of our chosen figure of merit with the initial entanglement negativity: the more entanglement we input at $t=0$ the higher the maximum of the entropy production rate is. We can get to the same conclusion (in absolute value) when we consider the negative peak $\Pi_{\rm min}$.

\begin{figure}
\centering
\includegraphics[width=0.83\linewidth]{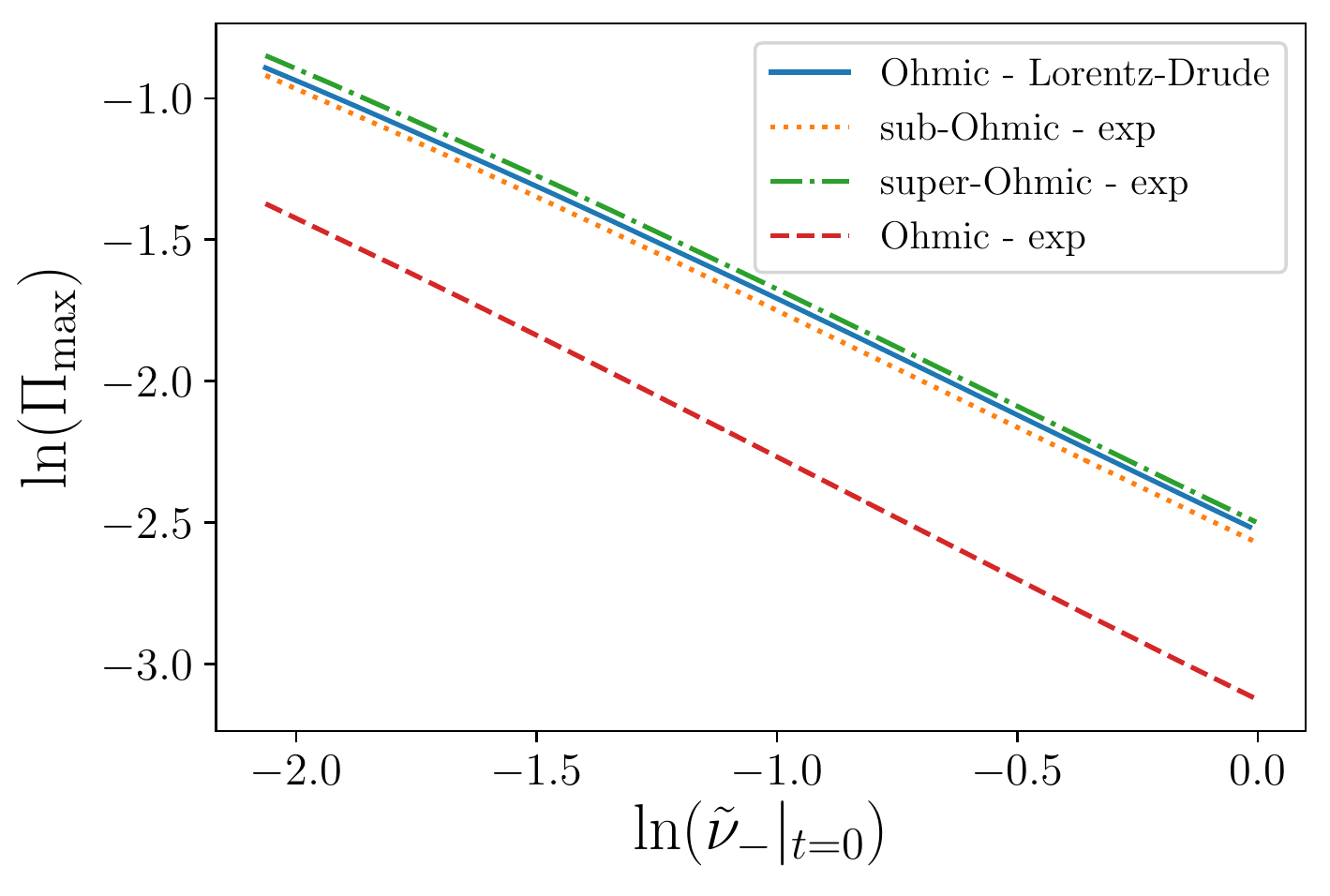}
\caption{\small{Plot of $\Pi_{\rm max}$ against the smallest symplectic eigenvalue of the partially transposed CM at $t=0$ (logarithmic scale) for different SDs (as stated in the legend). The initial state is prepared using the same parametrisation chosen in \Cref{fig5}, while we have taken $\alpha = 0.1 \omega_0$, $\omega_c = 0.1 \omega_0$, $\beta = 0.1 \omega_0^{-1}$.}}
\label{fig6}
\end{figure}

The monotonic behavior highlighted above holds regardless of the specific form of the spectral density.  In \Cref{fig6} we study $\Pi_{\rm max}$ against the smallest symplectic eigenvalue $\tilde{\nu}_{-}$ of the partially transposed CM for the various spectral densities we have considered, finding evidence of a power law of the form $\Pi_{\rm max} \propto \tilde{\nu}_{-}^{\delta}$, where the exponent $\delta$ depends on the reservoir's spectral properties.

\section{Markovian limit}
\label{sec:Markovian_Limit}
We are now interested in assessing whether the analytical and numerical results gathered so far bear dependence on the non-Markovian character of the dynamics. With this in mind, we explore the Markovian limit, in which the problem is fully amenable to an analytical solution, that can also be used to validate our numerical results. Such limit is obtained by simply choosing an Ohmic SD with a Lorentz-Drude regularisation -- \Cref{eq:SD_Ohm_LD} --  and taking the long time and high temperature limits, i.e. $\omega_0 t \gg 1$ and $\beta^{-1} \gg \omega_0$. This yields the time-independent coefficients
\begin{align}
\frac{\Delta(t) - \gamma(t)}{2}& \longrightarrow \gamma_M \left ( 2 \bar{n}(\omega_0) + 1\right ) , \\
\frac{\Delta(t) + \gamma(t)}{2}& \longrightarrow \gamma_M \, \bar{n}(\omega_0) \ ,
\end{align}
where $\bar{n}(\omega_0) = (e^{\beta \omega_0} -1)^{-1}$ is the average number of excitations at a given frequency $\omega_0$, whereas $\gamma_M \equiv 2 \alpha^2 \omega_c^2  \omega_0 / (\omega_c^2 + \omega_0^2)$. 
Therefore, \Cref{eq:ME_sec} reduces to a master equation describing the dynamics of two uncoupled harmonic oscillators undergoing Markovian dynamics, for which we take $\mathbf{A} = -\gamma_M \mathbbm{1}_4$ and $\mathbf{D}= 2 \gamma_M (2 \bar{n}(\omega_0) +1)  \mathbbm{1}_4$ in \Cref{eq:Lyapunov}.

\begin{figure}
\centering
\includegraphics[width=0.83\linewidth]{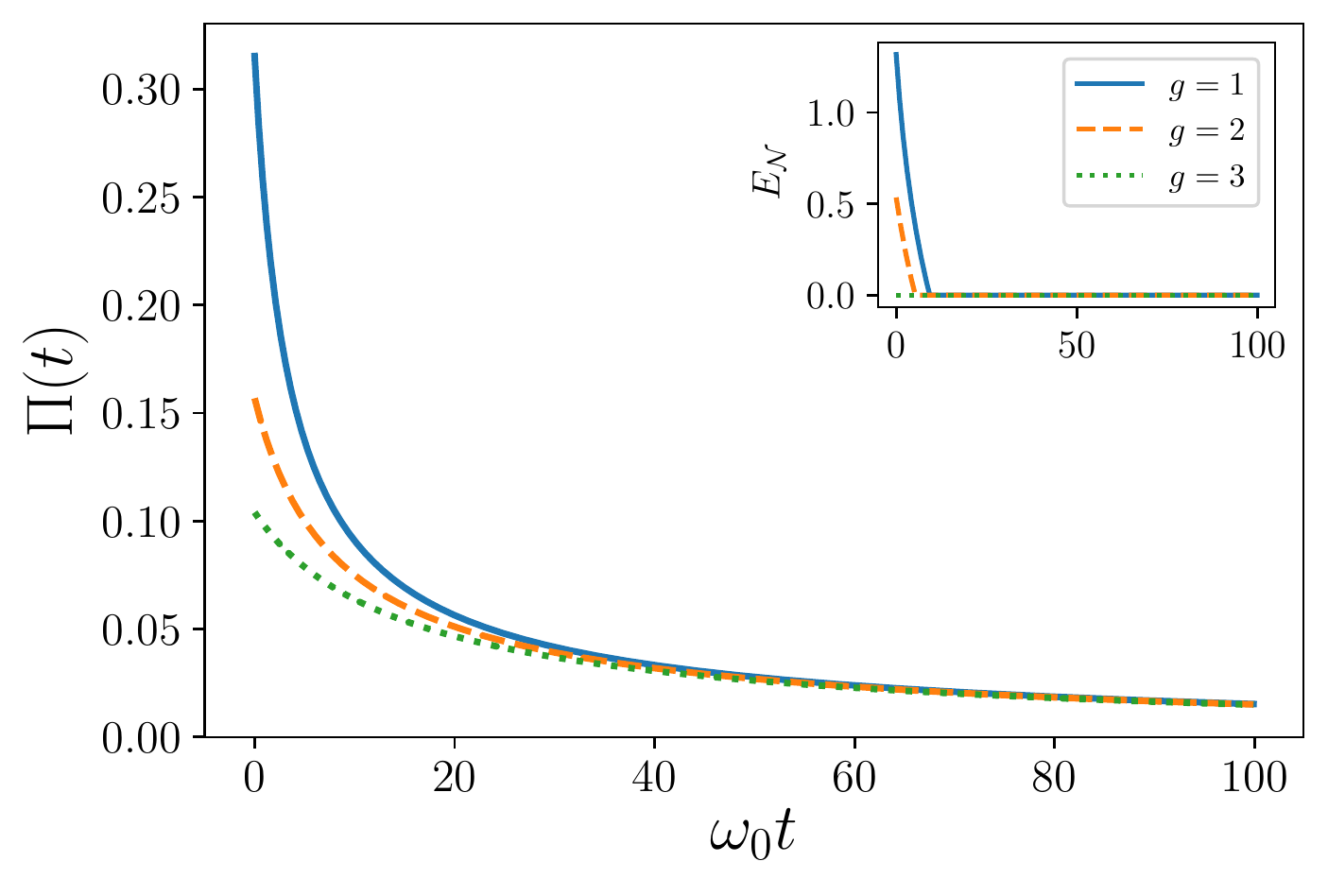}
\caption{\small{Entropy production rates corresponding to different preparations of the initial global state in the Markovian limit. We have taken different values of $g$ (thus varying the global purity of the state of the system) with $s=2$, $d=0$, $\lambda = 1$, $\alpha = 0.1 \omega_0$, $\omega_c = 0.1 \omega_0$, $\beta = 0.01 \omega_0^{-1}$.}}
\label{fig7}
\end{figure}

\begin{figure}
\centering
\includegraphics[width=0.83\linewidth]{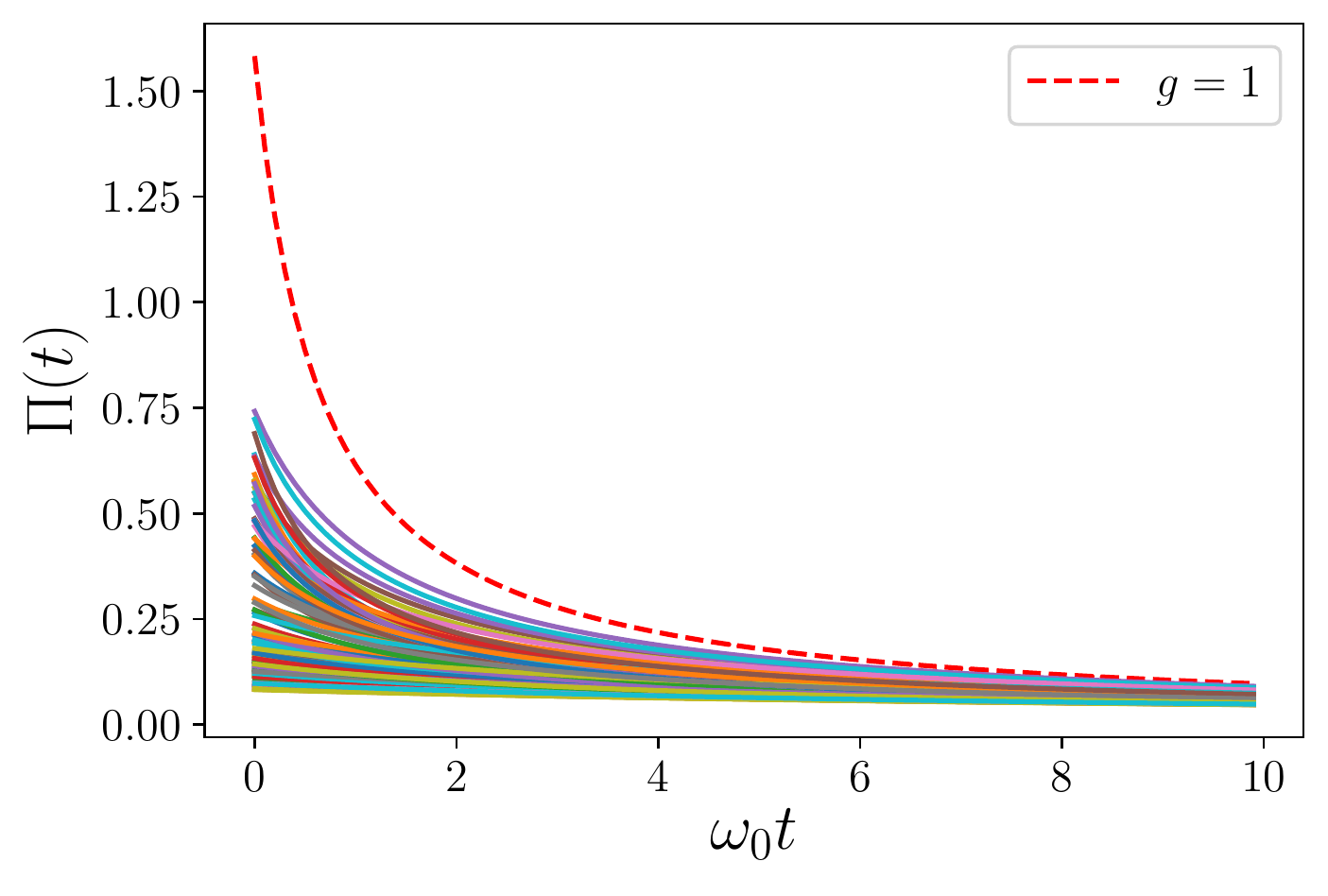}
\caption{\small{Entropy production rate $\Pi(t)$  plotted against the dimensionless time $\omega_0 t$ in the Markovian regime. The initial CM is parametrised by setting $s=10$ and randomly sampling (in a uniform manner) $d, g, \lambda$ from the intervals $[0,s-1]$, $[2d+1,d+10]$ and $[-1,1]$, respectively. We present $N_{R} = 100$ different realisations of the initial state. The dashed line represent the state with unit global purity ($g=1$) and $d=0, \; \lambda=1$. For the dynamics, we have taken $\alpha = 0.1 \omega_0$, $\omega_c = 0.1 \omega_0$, $\beta = 0.01 \omega_0^{-1}$.}}
\label{fig8}
\end{figure}

Working along the same lines  as in the non-Markovian case, we study the behavior of $\Pi(t)$ by suitably choosing the parameters encoding the preparation of the initial state. For example, in \Cref{fig7} we plot the entropy production rate as a function of time for different values of $g$. The limiting procedure gives back a coarse-grained dynamics monotonically decreasing toward the thermal state, to which it corresponds a non-negative entropy production rate, asymptotically vanishing in the limit $t \to \mathcal{1}$ . Moreover, the memoryless dynamics leads to a monotonic  decrease of the entanglement negativity, as shown in the inset of \Cref{fig7}.
In this case, the globally pure state ($g=1$, dashed line in \Cref{fig8})  still plays a special role: all the curves corresponding to value of $g$ smaller that the unity remains below it.

\begin{figure}
\centering
\includegraphics[width=0.83\linewidth]{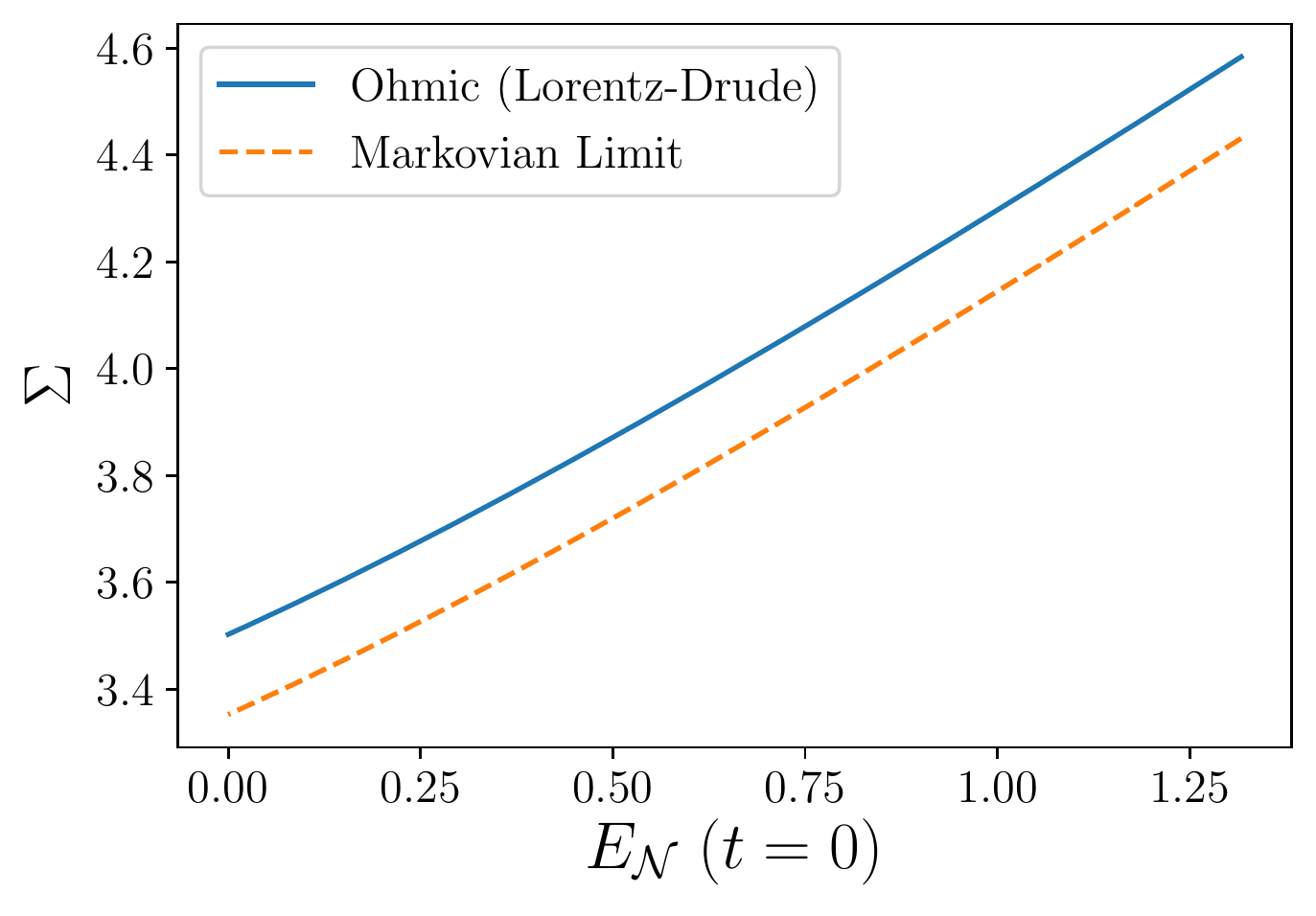}
\caption{\small{Entropy production in the non-Markovian case (solid line) as a function $E_{\cal N}(t=0)$, compared with its counterpart achieved in the corresponding Markovian limit (dashed line).  We have taken $s=2$, $d=0$, $\lambda = 1$, $\alpha = 0.1 \omega_0$, $\omega_c = 0.1 \omega_0$, $\beta = 0.01 \omega_0^{-1}$.}}
\label{fig9}
\end{figure}

The Markovian limit provides a useful comparison in terms of integrated quantities. In this respect, we can study what happens to the entropy production $\Sigma = \int_{0}^{+\mathcal{1}} \Pi(t) dt$. Although the non-Markovian dynamics entails the negativity of the entropy production rate in certain intervals of time, the overall entropy production is larger than the quantity we would get in the corresponding Markovian case, as can be noticed in \Cref{fig9}. 

Finally, we can study the dependence of the Markovian entropy production rate on the initial entanglement. Note that, in this limit, \Cref{eq:entropy_prod_rate} yields an analytic expression for the entropy production rate at a generic time $t$, which we write explicitly in \Cref{eq:entropy_prod_rate_Markov}. From our numerical inspection, we have seen that the entropy production rate is maximum at $t=0$, so that
\begin{align}
\label{eq:entropy_prod_rate_Markov_max}
\Pi_{\rm max} \equiv \Pi(0) & = - 8 \gamma_M + 4 s \, \gamma_M \tanh\left (\frac{\beta \omega_0}{2} \right ) \nonumber \\ 
& +\frac{4 s \, \gamma_M \coth \left (\frac{\beta \omega_0}{2} \right )}{ (2 s -\tilde{\nu}_{-}) \tilde{\nu}_{-}} \ .
\end{align} 

\begin{figure}
\centering
\includegraphics[width=0.83\linewidth]{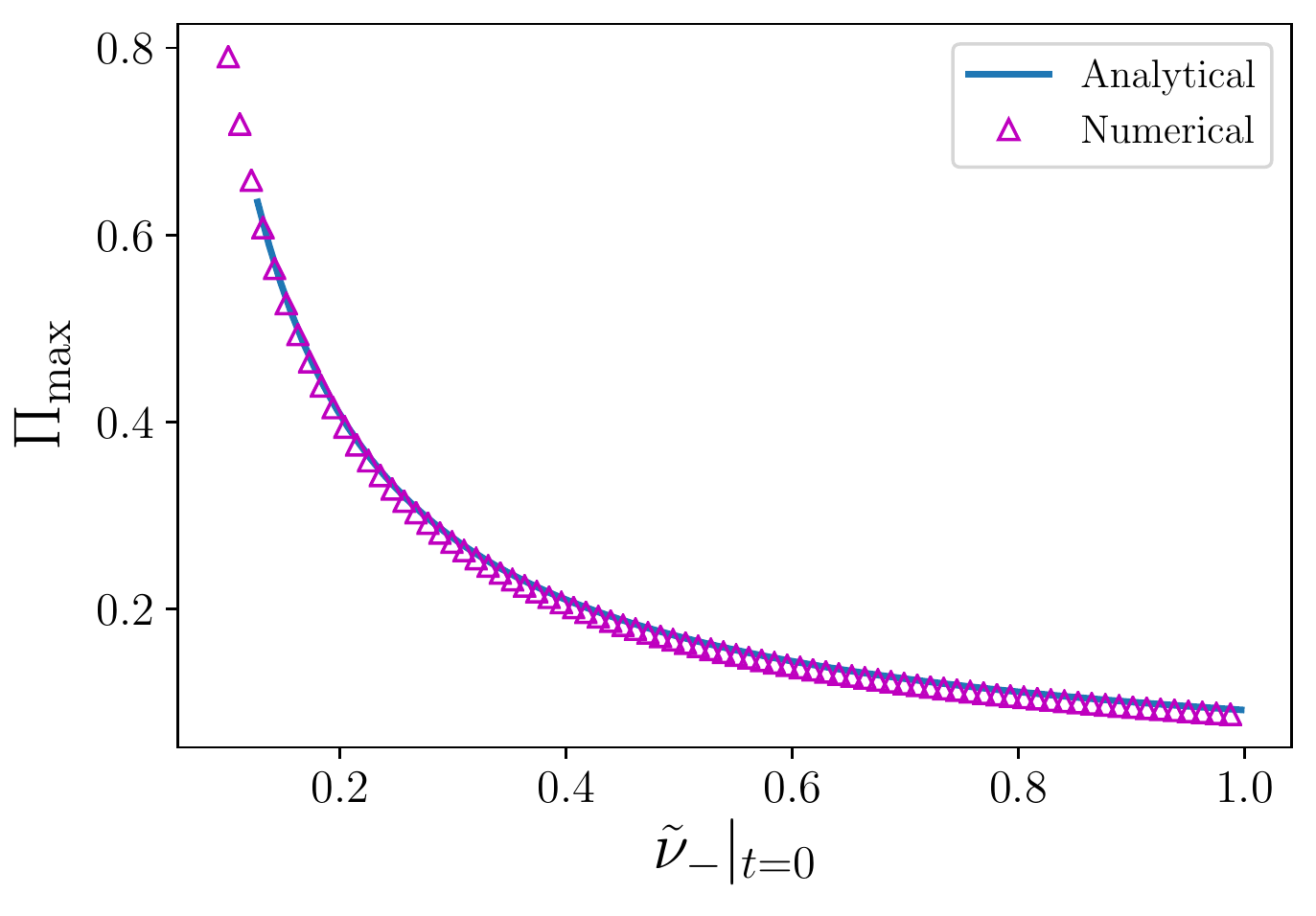}
\caption{\small{Markovian limit: maximum entropy production rate $\Pi_{\rm max}$  as a function of the minimum symplectic eigenvalue of the partially transposed CM at $t=0$. We have taken $s=4$, $g=2d +1$, $\lambda = 1$, $0 \le d \le 3$, $\alpha = 0.1 \omega_0$, $\omega_c = 0.1 \omega_0$, $\beta = 0.1 \omega_0^{-1}$. We compare the curve obtained numerically (triangles) to the analytical trend (solid line) found through \Cref{eq:entropy_prod_rate_Markov_max}.}}
\label{fig10}
\end{figure}

If we fix the parameter $s$ and plot $\Pi_{\rm max}$ against $\tilde{\nu}_{-}$, we can contrast analytical and numerical results [cf.~\Cref{fig10}]. We can draw the same conclusion as in the non-Markovian case: the more entanglement we input, the higher the entropy production rate.

\section{Conclusions}
\label{sec:conclusions}
We have studied -- both numerically and analytically -- the dependence of the entropy production rate on the initial correlations between the components of a composite system. We have considered two non-interacting oscillators exposed to the effects of local thermal reservoirs. By using a general parametrisation of the initial state of the system, we have systematically explored different physical scenarios. We have established that correlations play an important role in the rate at which entropy is intrinsically produced during the process. Indeed, we have shown that, when the system is prepared in a globally pure state,  we should expect a higher entropy production rate. This is the case -- regardless of the spectral density chosen -- for initial entangled states of the oscillators: larger initial entanglement is associated with higher rates of entropy production, which turns out to be a monotonic function of the initial degree of entanglement. 
Remarkably, our analysis takes into full consideration signatures of non-Markovianity in the open system dynamics.  

It would be interesting, and indeed very important, to study how such conclusions are affected by the possible interaction between the constituents of our system, a situation that is currently at the focus of our investigations, as well as non-Gaussian scenarios involving either non-quadratic Hamiltonians or spin-like systems.

\acknowledgements
We thank R. Puebla for insightful discussions and valuable feedback about the work presented in this paper.
We acknowledge support from the H2020 Marie Sk{\l}odowska-Curie COFUND project SPaRK (Grant nr.~754507), the H2020-FETPROACT-2016 HOT (Grant nr.~732894), the H2020-FETOPEN-2018-2020 project TEQ (Grant nr.~766900), the DfE-SFI Investigator Programme (Grant 15/IA/2864), COST Action CA15220, the Royal Society Wolfson Research Fellowship (RSWF\textbackslash R3\textbackslash183013) and International Exchanges Programme (IEC\textbackslash R2\textbackslash192220), the Leverhulme Trust Research Project Grant (Grant nr.~RGP-2018-266). 

\appendix

\section{Analytic expressions for the entropy production rate}\label{app:a}

We report the expression for the entropy production rate in the non-Markovian regime and the Markovian limit. By using $\mathbf{A} = -\gamma(t) \mathbbm{1}_4$, $\mathbf{D}= 2 \Delta(t)  \mathbbm{1}_4$ and $\boldsymbol{\sigma} (t)$ given by \Cref{eq:sigma_t_wc}, one obtains the following expression for the general non-Markovian case
\begin{align}
\label{eq:entropy_prod_rate_general}
& \Pi(t) =  - 8 \gamma(t) + \frac{4\gamma^2(t)  \left (s - 2 s \; \Gamma(t) + \bar{\Delta}(t)\right )}{\Delta(t)} \nonumber \\ & + \frac{4 \Delta(t) \left (s - 2s \; \Gamma(t) +\bar{\Delta}(t) \right )}{\tilde{\nu}_{-}(2s - \tilde{\nu}_{-} ) \left ( 1 -  \Gamma(t) \right)^2 + 2 s  \bar{\Delta}(t) \left ( 1 -  \Gamma(t) \right) +  \bar{\Delta}^2(t)  }, 
\end{align} 
where $ \bar{\Delta}(t) = 2 \int_0^{t} \Delta(\tau) d \tau$.

On the other hand, in the Markovian limit discussed in Sec. \Cref{sec:Markovian_Limit}, \Cref{eq:entropy_prod_rate_general} reads
\begin{widetext}
\begin{align}
\label{eq:entropy_prod_rate_Markov}
\Pi(t) = - 8 \gamma_M + 4 \gamma_M \left [  1 + e^{- 2 \gamma_M t}  s \tanh\left (\frac{\beta \omega_0}{2} \right )\right ] 
+ \frac{ 4 \gamma_M  \coth \left ( \frac{\beta \omega_0}{2}\right ) e^{2 \gamma_M t}\left[ s+ \left ( e^{2 \gamma_Mt} - 1\right ) \coth \left ( \frac{\beta \omega_0}{2}\right ) \right ]}{ \left [ 2s -\tilde{\nu}_{-}+ \left ( e^{2 \gamma_Mt} - 1\right ) \coth \left ( \frac{\beta \omega_0}{2}\right )  \right ]  \left [ \tilde{\nu}_{-}+ \left ( e^{2 \gamma_Mt} - 1\right ) \coth \left ( \frac{\beta \omega_0}{2}\right )  \right ]} \ .
\end{align}
\end{widetext}

\bibliographystyle{apsrev4-1.bst}
\bibliography{Entropy_Prod.bib}

\end{document}